\documentclass[epj,referee]{svjour}
\usepackage{graphicx}
\begin{document}
\title{Models for the size distribution of businesses in a price driven market}
\author{R. D'Hulst \and G. J. Rodgers
}                     
%
%
\institute{Department of Mathematical Sciences, Brunel University, Uxbridge, Middlesex UB8 3PH, UK}
\date{Received: date / Revised version: date}
%
\abstract{
A microscopic model of aggregation and fragmentation is introduced to investigate the size distribution of businesses. In the model, businesses are constrained to comply with the market price, as expected by the customers, while customers can only buy at the prices offered by the businesses. We show numerically and analytically that the size distribution scales like a power-law. A mean-field version of our model is also introduced and we determine for which value of the parameters the mean-field model agrees with the microscopic model. We discuss to what extent our simple model and its results compare with empirical data on company sizes in the U.S. and debt sizes in Japan. Finally, possible extensions of the mean-field model are discussed, to cope with other empirical data.
} 
\maketitle
\section{Introduction}
\label{sec:introduction}

In simple terms, the life of a business can be thought of as three phases, creation, growth and death. There are numerous motivations for the creation of a new business, but the main principle is that it must attract enough customers to be profitable. The actual number of customers of a business is of course strongly time dependent, as the British retail giant M\&S discovered \cite{marks_and_spencer}, and some very popular businesses can become unpopular in a matter of months. Finally, there is little doubt that the fate of every business, sooner or later, is to close, normally for lack of popularity or bankruptcy. The history of each business is of course particular, its creation, successes and failures all having their own reasons. However, all businesses are part of the larger industry of selling to customers and they are in competition with one another to survive. This interdependence between businesses is likely to induce group behaviour. This group dynamics applies for shops or companies, where lack of popularity or bankruptcy are common fates. A good example of the former is Smith and Corona, once the largest typewriter maker in the US, which was supplanted by the electronic revolution and had to close this year after a 114 year history \cite{smith_corona}, while the latter is best illustrated by the fate of Boo.com \cite{boo_com}, an internet clothing retail company that spent \$135 million of its investors money before having to close down. In this work, we investigate the size distribution of businesses in a model of fragmentation and aggregation.

We define a very simple model where customers try to find a business that fits their expectations as well as possible, gathering information at random. The businesses themselves are constrained by the fact that most customers are 
expecting a particular price, and businesses that do not offer this price are likely to go bankrupt. No attempt is made to mimic any real situation, we have simplified the model as much as possible in an attempt to capture the salient features of the process. For instance, a customer either goes to one business or none, and all businesses are completely identical, that is, the only difference between businesses is the price they offer. Both these assumptions are highly unrealistic and we discuss in Sec. \ref{sec:comparison with empirical data} to what extent the results of our model compare with empirical data. The ultimate purpose of our model is to address the dynamics of the cooperative behaviour of selling organisations interacting through the customers that they try to attract. As such, we expect our model to point towards universal features of organisations that are competing to attract customers, like shops or companies. To avoid any confusion, we will stick to the generic terms business to denote the selling organisation and customer for a buying agent, postponing our comparison with real life to Sec. \ref{sec:comparison with empirical data}. This will allow us to discuss the assumptions of our model in the light of a comparison between its results and empirical data. 

The model is defined in detail in Sec. \ref{sec:the model}, previous models are recalled in Sec. \ref{sec:previous models}, numerical simulations are presented in Sec. \ref{sec:numerical results} and they are analysed in a mean-field framework in Sec. \ref{sec:analytical results}. Our results are compared with empirical data in Sec. \ref{sec:comparison with empirical data} and summarised in the last section.

\section{The model}
\label{sec:the model}

The model is made up of customers and businesses, where a business corresponds to a cluster of customers. A customer can only go to one business or none. Both customers and businesses are given prices. The price associated with a customer is an estimate of the amount of money he can spend. Initially, these prices are chosen from a probability distribution $D (p)$. We denote the average price over all customers by $\overline{p}$. The price of a business is the price paid by all its customers, that is, all customers of a business are associated to the same price, which is the price of the business. This price is fixed by the first customer which, because of this special role, can be seen as the owner of the business. At the beginning of the simulation, there are only clusters of one customer. Hereafter, we consider that the number of customers, $N_0$, is fixed.

At each time step, a customer $i$ is chosen at random. If $i$ belongs to a business, this business goes bankrupt with probability $\alpha_i = | p_i - \overline{p} |$, or nothing happens with probability $1 - \alpha_i$. Remember that $p_i$ is also the price of the business. This assumes that a business that is badly adapted to the commonly agreed price cannot survive. The bankruptcy originates from the desertion of customers from high prices or from financial slump for too low prices. In economics, this refers to the theory of perfect competition, which implies that the market price is equal to the marginal cost, preventing a firm from just choosing a price. When a business goes bankrupt, all its customers are given new expected prices in a range $r$ around their previous common price $p$. If the chosen customer $i$ is a cluster of size one, he is a free customer who wants to make up his mind upon which business he prefers. Another customer $j$ is chosen at random and with a probability $| p_i - p_j |$, $i$ decides to go to the same business as $j$, otherwise, he remains a free customer. This process suggests that customers with similar expectations are more likely to go to the same business. Then, $i$ is forced to comply with the price proposed by the business and his expected price is changed to $p_j$. We have not excluded the possibility that $j$ could be a cluster of size one. Hence, if $i$ is a cluster of size one, he is a free customer, while if $j$ is a cluster of size one, he is the first customer or the owner of a new business. 

In this model, both businesses and customers are constrained. Businesses have to comply with the market agreed price, otherwise they go bankrupt, while customers can only choose between existing businesses. We would say that customers are price takers. These two processes can be compared to a democratic and a dictatorship constraints respectively, as defined in previous models of herding \cite{dhulst00}. It was shown in \cite{dhulst00} that democratic aggregations lead to a convergence towards one common value for $p$, which will be identified as the market price here. Dictatorship aggregations lead to a spontaneous segregation in the population, but as shown in \cite{dhulst-apfa2}, a mixed version of both aggregation processes tends to favor the democratic convergence. 

As can be appreciated, the model is very simple and is based on the assumption that the market is essentially price driven. We assume that customers are only going to one particular business, all businesses being identical. No spatial structure is introduced, customers do not have any notion of a business being closer than another. Finally, customers only consider going to another business if their business is going bankrupt. All these assumptions are likely to have a different impact on different applications of our model, and they are best discussed in Sec. \ref{sec:comparison with empirical data}, where we compare the predictions of our model to different empirical data. Finally, let us also mention that we only consider the model in its relation with customers considering different prices, but that it could be reformulated in many different situations. We could speak of investors instead of customers and  consider that the $p_i$'s are interest rates instead of prices, for instance.

\section{Previous models}
\label{sec:previous models}

The size distribution of businesses is not a new field of research and it has already attracted a considerable amount of work in economics, and more recently in physics. We delay to Sec. \ref{sec:comparison with empirical data} the presentation of some empirical studies, to be able to compare with results of the model, while we consider here previous models for the growth of businesses.

Pareto \cite{pareto}, Gibrat \cite{gibrat} and Zipf \cite{zipf} among others, have shown that power-laws are found in many different areas of sciences, such as size distribution of taxinomic genera, word frequency in modern languages or business size distribution, for examples. As these and many other subjects are only remotely related to each other, it is obvious that the similarities can only arise from the underlying dynamics, which should be independent of the details of the particular system considered. Champernowne was one of the first to consider a stochastic model for the growth of businesses \cite{champerone}, and he already noticed that there are so many factors influencing a business, that any model should either be unrealistically simplified or hopelessly complicated. He went for the first option and proposed a model where businesses belong to a given income range, and can move from one income range to another, the total wealth being conserved. Simon proposed a similar model, that he introduced in the context of word frequency \cite{simon}, and extended to reproduce income distributions. Later, Simon and Bonini made a review of the subject, emphasizing the need for further research into the processes that generate such distributions \cite{simon2}. 

A major breakthrough in this direction was then made by Mandelbrot \cite{mandelbrot-levy}, with his seminal works on L\'evy distributions. His original work was concerned with price variations, but it has been extended since then. The main result is that the distribution of a sum of random variables does not necessarily converge towards a Gaussian distribution, but it can also converge towards a L\'evy distribution. These distributions are particularly suitable to describe large deviations, having power-law tails, with an exponent that varies between 1 and 3. It is interesting to note that in a very complete review of the growth of firms \cite{steindl}, Steindl suggested that the failure to converge towards a Gaussian distribution can be attributed to boundary or other considerations, while the author was not convinced that the alternative explanation put forward by Mandelbrot would be helpful in economics. This illustrates the complete change of perception that we have at the present time.

More recently, physicists have been considering models of company size distributions, based on the internal structure of a company \cite{stanley-model}, or on the exchange of capital between competing firms \cite{sid2,takayasu98}. One of the foci of these models has been the results for firm growth from Ref. \cite{amaral97}.

The model proposed here is very similar to the models of Champernowne \cite{champerone} or Simon \cite{simon}. To solve it, a mean-field model is introduced, replacing the microscopic price dynamics by two macroscopic parameters of aggregation and fragmentation. Allowing one of these parameters to depend on a business size, we are able to determine the size distribution of the clusters of customers. The particular case of business distributions is interesting as it requires a particular tuning to obtain the required power-law distribution. The introduction of the macroscopic model also opens the way for further research into different size distributions and, as is shown in Ref. \cite{nato2001}, more general distributions can be generated. This work illustrates the interaction between microscopic and macroscopic models. The microscopic price dynamics is first introduced, motivating the macroscopic model, which can be made more general and motivates further research for the corresponding microscopic models. 

\section{Numerical results}
\label{sec:numerical results}

\begin{figure}
\includegraphics[width=0.5\textwidth]{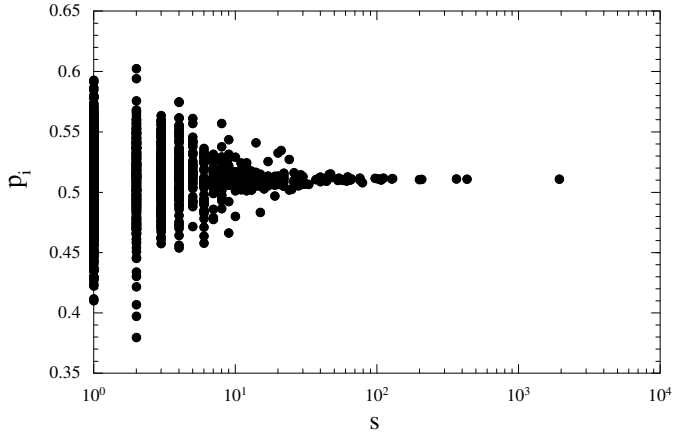}
\caption{Price $p_i$ of a business as a function of its number of customers, for $N_0 = 10^4$ customers and a range $r = 0.1$. The result is taken at the end of a simulation of $10^6$ time steps.}
\label{fig:price(size) for businesses}
\end{figure}
We have performed numerical simulations of the model for a population of $10^4$ customers and an initially uniform distribution $D (p)$. In Fig. \ref{fig:price(size) for businesses}, we report the prices $p_i$ of the businesses as a function of their number of customers for one simulation of $10^6$ time steps with a range $r$ equal to 0.1. The very large maximum around $p \approx 0.5$ shows that only businesses that are offering a price close to the average price over all customers can grow. We call the average price $\overline{p}$ the market price. Going away from this maximum, businesses with an exponentially decreasing size can form, with a steeper decrease as $r$ gets smaller. Fig. \ref{fig:price(size) for businesses} is not difficult to understand if we consider the following equation for the size $s$ of a business proposing a price $p$ at time $t$,

\begin{eqnarray}
\nonumber
E_t (s (p,t+1) - s (p,t)) &=& \frac{s n_1 (1 - \langle |p - p_j (1) |\rangle_1)}{N_0}\\
&-& \frac{s (s -1)}{N_0} |p - \overline{p}|,
\label{eq:size as a function of the price}
\end{eqnarray}
using $E_t (x)$ for the expectation of $x$ at time $t$. $p_j (s)$ is defined as the price proposed by the $j$\raisebox{0.5mm}[0mm][0mm]{\footnotesize th} business of size $s$ and $n_1$ is the number of clusters of size 1 per customer. $\langle |p - p_j (1) |\rangle_1$ is the average value of $| p - p_j (1)| $ over all clusters of size 1. The first term on the right hand side of Eq. (\ref{eq:size as a function of the price}) describes the growth of a business of size $s$ proposing a price $p$, while the second term takes into account the probability of bankruptcy of this business. If we assume that the system is infinite and in its stationary state, $n_1$, $\overline{p}$ and $\langle |p - p_j (1) |\rangle_1$ are all time independent, and we can solve for $s (p, t)$ in the limit $t \rightarrow \infty$ to obtain

\begin{equation}
s (p) = 1 + n_1 \frac{1 - \langle |p - p_j (1) |\rangle_1}{|p - \overline{p}|}.
\label{eq:s as a function of p}
\end{equation}
$s (p)$ corresponds to the average size that a business proposing a price $p$ reaches. Of course, $s (p) \ge 1$ and diverges for $p = \overline{p}$, as can be seen in Fig. \ref{fig:price(size) for businesses}, where most businesses are proposing a price close to $\overline{p}$.

The value of the market price $\overline{p}$ is approximately equal to the first moment of the initial distribution $D (p)$. Superimposing the results of Fig. \ref{fig:price(size) for businesses} for several simulations gives a distribution with a broad maximum extending from $0.5-r/2$ to $0.5+r/2$, for an initially uniform distribution, with exponentially decreasing tails out of this range. The existence of a commonly agreed market price is the relevant result, while the exact value of this price is meaningless. Variants of the model can be devised to make the market price converge towards different values by changing the way a customer picks his business. For instance, $p_i$ could be considered as the maximum amount of money available to customer $i$. For this variant and others, the same analysis can be carried out by changing the expression $1 - \langle |p - p_j (1) |\rangle_1$ according to the details of the variant. The main conclusion remains unchanged, namely that $s (p)$ diverges at $\overline{p}$, so that most businesses are offering the same price. Also, one could choose to fix a parameter instead of taking $\overline{p}$ as reference value, without changing the results. This, however, would be much more artificial.

\begin{figure}
\includegraphics[width=0.5\textwidth]{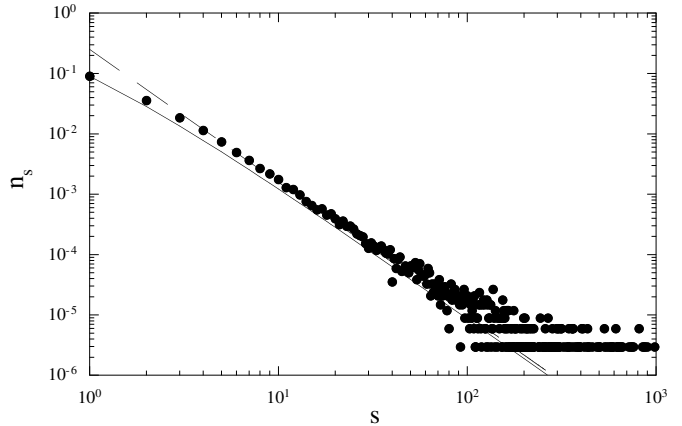}
\caption{Cluster size distribution for the businesses ($\bullet$). The continuous line is the size distribution given by Eq. (\ref{eq:n_s for beta = 1}) for $\alpha = 1.20$. The dashed line is a guide to the eye for a power-law of exponent $\tau = 2.20$, the expected asymptotic behaviour.}
\label{fig:cluster size distribution}
\end{figure}
In Fig. \ref{fig:cluster size distribution}, we present the size distribution $n (s)$ of businesses of size $s$ for the same values of the parameters.  A power-law $n (s) \sim s^{-\tau}$ with an exponent $\tau = 2.20 \pm 0.05$ can be seen for large $s$. This compares with the empirical findings of Nagel {\it et al} \cite{nagel00} for the US establishment and firm sizes, as estimated from annual sales in the retail sector. In this study, the authors found that the size distributions are power-laws of exponent $\tau = 2$. But as already explained, we postpone to Sec. \ref{sec:comparison with empirical data} a detailed comparison of the model with empirical data.

\section{Analytical results}
\label{sec:analytical results}

To investigate this model analytically, it is convenient to consider a mean-field version. At each time step, a customer $i$ is chosen at random. If he belongs to a cluster of size $s >1$, the business he is going to goes bankrupt with a probability $a$. That is, his cluster of $s$ customers is transformed into $s$ independent clusters of one customer with a probability $a$. If $i$ is a cluster of size one, another customer $j$ is chosen at random and $i$ becomes a customer of the same business as $j$ with a probability $b$. The microscopic dynamics generated by the prices is replaced by two macroscopic parameters $a$ and $b$. As in the present model all businesses are alike apart from their number of customers, we assume that these parameters $a$ and $b$ can be functions of the size of a business only. Moreover, the parameter $b$ expresses the time scale for the information transmission between different customers. For simplicity, we assume that customers are gathering information at random, which means that they are more likely to gather information about businesses with a large number of customers than small businesses. Hence, we assume that $b$ is a constant and fix its value to 1. The validity of this assumption is discussed somewhere else \cite{nato2001}. It implies that the difference in time scales between bankruptcies and customer choices is controled by $a (s)$. The relation between the microscopic prices and the macroscopic function $a (s)$ is

\begin{equation}
a (s) = \frac{1}{n_s} \sum_{j=1}^{n_s} | p_j (s) - \overline{p} |
\label{eq:a as micro -> macro}
\end{equation}
where the summation is running over all clusters of size $s$, and $p_j (s)$ is the price proposed by the $j$\raisebox{0.5mm}[0mm][0mm]{\footnotesize th} business of size $s$, as previously defined.

The master equations for the evolution of the number $n_s$ of businesses of size $s$ per customer at time $t$ for $s>1$ can be written as

\begin{equation}
\frac{\partial n_s}{\partial t}Ê= - a (s) s n_s \left( 1 + \frac{n_1}{a (s)} \right) + n_1 (s-1) n_{s-1}.
\label{eq:master equation n_s}
\end{equation}
Note that one time step in this continuous description is chosen to correspond to one attempted update per agent in the numerical simulation. The first term on the right hand side represents the disappearance of businesses of size $s$, either because they go bankrupt, or because they increase their size to $s+1$. The second term describes the appearance of new businesses of size $s$ due to businesses of size $s-1$ attracting a new customer. For $s=1$, the relation is

\begin{equation}
\frac{\partial n_1}{\partial t} = \sum_{s=2}^{\infty} a (s) s^2 n_s - n_1 \left( \sum_{s=1}^{\infty} s n_s + n_1\right).
\label{eq:master equation n_1}
\end{equation}
The first term on the right hand side corresponds to the appearance of $s$ new independent customers because of the bankruptcy of a business of size $s$, which happens with a probability $a (s)$. The second term describes the disappearance of independent customers because they choose to go to a particular business. Note that $n_1$ appears twice in this second term because of the two possible natures of clusters of size one.

As customers are chosen at random, larger businesses are more often selected than smaller ones. Hence, if the bankruptcy probability of a selected business is equal to $a (s)$, its effective probability of going bankrupt is equal to $a_{eff} = s a (s)$. In the following, we consider that $a (s) = \alpha n_1 s^{- \beta}$, where $\alpha$ and $\beta$ are free parameters. The multiplicative factor $n_1$ is introduced for convenience. The value of $\beta$ determines whether bankruptcies are more likely to affect the small or large businesses. For $\beta < 1$, the bigger the business, the larger the effective probability of going bankrupt, and conversely for $\beta > 1$. $\beta = 1$ gives an effective bankruptcy probability $a_{eff}$ independent of the businesses size. For actual situations, we expect $\beta$ to be greater or equal to one, as bigger businesses are usually firmly established. We call the parameter $\alpha$ the rate of bankruptcy because it is a measure of the difference in time scales between bankruptcies and customers' choices. Note that as $\alpha$ is multiplied by $s^{-\beta}$, it does not necessarily have to be small to have a small probability of going bankrupt. 

Even if it seems rather arbitrary to take $a (s)$ as a power-law, this choice is, at least, not unreasonable. First, it has $a (s)$ independent of $s$ as special case, which is particularly suitable as empirical data on company bankruptcy show little dependence on size \cite{amaral97}. Second, as will be shown, it can reproduce the results of the proposed microscopic model. Third, as noted in early work by Pareto \cite{pareto}, Gibrat \cite{gibrat} and Zipf \cite{zipf} amongst many others, power-laws are ubiquitous in nature, which means that taking $a (s)$ as a power-law is a reasonable starting point. However, one should notice that this is not the only possibility and that other possibilities are investigated elsewhere \cite{nato2001}.

The stationary solution of Eq. (\ref{eq:master equation n_s}) is given by

\begin{equation}
(\alpha s^{-\beta} + 1 ) s n_s = (s-1) n_{s-1}
\label{eq:iterative relation for sn_s}
\end{equation}
which has the formal solution

\begin{equation}
n_s = \frac{n_1}{s} \prod_{r=2}^{s} \frac{1}{1 + \alpha r^{-\beta}}.
\label{eq:solution for s n_s}
\end{equation}
A similar expression was found for the average number of sites with a given number of links in growing random networks \cite{sid_and_paul}. For large $s$, this expression has the asymptotic behaviour

\begin{equation}
n_s \sim \left\{
\begin{array}{ll}
s^{-1 - \alpha}& \beta = 1,\\
s^{-1} \exp \left[ - \alpha \left( \frac{s^{1-\beta}- 2^{1-\beta}}{1 - \beta} \right) \right]& \frac{1}{2} <\beta < 1,\\
s^{\frac{\alpha^2}{2}-1} \exp \left[- 2\alpha \sqrt{s} \right]& \beta = \frac{1}{2},\\
s^{-1} \exp \left[ - \alpha \frac{s^{1-\beta}}{1 - \beta} + \frac{\alpha^2}{2}\frac{s^{1-2\beta}}{1-2\beta} \right]& \frac{1}{3} <\beta < \frac{1}{2},\\
\end{array}
\right.
\end{equation}
and so on. As in \cite{sid_and_paul}, whenever $\beta$ decreases below $1/m$, with $m$ a positive integer, an additional term in the exponential becomes relevant. For values of $\beta$ greater than 1, the model does not have any stationary state because the fragmentation rate cannot compensate the aggregation process. This corresponds in any finite system to a monopolistic market, where one business attracts all customers. 

The numerical results of the microscopic model suggest that $\beta = 1$ and that $\alpha$ is close to but bigger than 1. This can be justified by combining Eqs. (\ref{eq:s as a function of p}), which implies that $s \sim |p - \overline{p}|^{-1}$ and (\ref{eq:a as micro -> macro}) with $a (s) = \alpha n_1 s^{- \beta}$. This gives $\alpha s^{-\beta} \sim s^{-1}$, or $\beta = 1$, a result that we verified numerically. For $\beta = 1$, the size distribution is equal to

\begin{equation}
n_s = \frac{(s-1)! (\alpha + 1)!}{(\alpha + s)!}n_1.
\label{eq:n_s for beta = 1}
\end{equation}
The continuous line presented in Fig. \ref{fig:cluster size distribution} corresponds to this equation for $\alpha = 1.20$. This value of $\alpha$ has been determined numerically using a power-law fit for large $s$, the value of $n_1$ and the value of the number of businesses. From these three different measurements, we obtained that $\alpha = 1.20 \pm 0.05$. The relation between $n_1$, the number of businesses and $\alpha$ is given below. 
 
To determine the value of $\alpha$ in the microscopic model, we first calculate $n_1$ from the stationary solution of Eq. (\ref{eq:master equation n_1}), which gives 

\begin{equation}
n_1 = \frac{\alpha -1 }{\alpha + 1}
\end{equation}
for $\beta = 1$. This implies that $\alpha$ should be greater than or equal to one to have a sensible solution. In effect, in any real situation, we expect $a_{eff} = s a(s)$ to be small because the process of choosing a business is much faster than the bankruptcy rate. Setting $a_{eff} = \epsilon$, where $\epsilon$ is a small parameter, and expanding to the first order in $\epsilon$, it is easy to show that $\alpha = 1 + 2\epsilon$ and that $\epsilon \sim 1/N_0$. Otherwise stated, $\alpha$ is greater than but close to 1, converging to 1 as the size of the system diverges.

In both models, for a customer to decide to join a business, another customer of this business has first to be selected. If we consider that a link is created between these two customers, the connectivity $c$ of this network, defined as the average number of links per customer, is related to the number $M_0$ of businesses per customer according to

\begin{equation}
c = 2 \left( 1 - \frac{M_0}{M_1}\right)
\label{eq:connectivity}
\end{equation}
because a business of size $s$ has $s-1$ links, or $2 (s-1)$ links for $s$ customers. In Eq. (\ref{eq:connectivity}), $M_k$ denotes the moment of order $k$ of the distribution $n_s$. Of course, $M_0$ is the number of businesses per customer and $M_1 = 1$, for normalization. The detailed structure of the network is very simple, as businesses are not connected with each other, and the $2 (s-1)$ links per $s$ customers of a business of size $s$ are connected at random, with at least one link starting from each customer. Using Eq. (\ref{eq:iterative relation for sn_s}) for $\beta = 1$, the number of businesses per customer can be calculated and we find $M_0 = (\alpha - 1)/\alpha$. As $n_s \sim s^{-1-\alpha}$, all moments of order $k \ge 1+ \alpha$ are diverging and because $\alpha$ is close to one for the microscopic model, only $M_0$ stays finite. 

Finally, the variations of the market price generated by the microscopic model can be investigated using our results for the size distribution. The large changes in the market price originate from bankruptcies, when a consequent number of customers are given new expected prices in a given range $r$ around their old price. When a business of size $s$ goes bankrupt, which happens with a probability $s a (s) n_s$, the average change in the market price is of the order of $r\sqrt{s}/4$. So, if a return $R$ is defined to be the change in the market price in one time step, the probability $P (R)$ of having a return of size $R$ scales like $P (R) \sim 1/R$ for large $R$. We checked that this result agrees with our numerical simulations.

\section{Comparison with empirical data}
\label{sec:comparison with empirical data}

Throughout this work, we have been using the term business to refer to an organisation that sells a product or a commodity. After having analysed the models, we can now consider their range of applicability. There has been a lot of work investigating empirically the size distribution of different selling organisations. One of the first problems was to choose a relevant quantity to describe the size of a selling organisation, but as is explained in \cite{amaral97} for instance, using sales, number of employees or market capitalisation gives essentially the same result. This allows us to discuss results borrowed from different sources of empirical data. In 1897, Pareto \cite{pareto} proposed a power-law distribution for the wealth of individuals. More than 100 years later, this result is still a topical question, and we will concentrate on recent empirical data. Nagel {\it et al} \cite{nagel00} reported results for U.S. establishment size and firm size in the retail sector, using annual sales as a proxy for the size. They reported that both size distributions display a maximum for a typical annual size of approximately $\$400, 000$. Around this typical size, both distributions are lognormal. However, for larger annual sales, a power-law like

\begin{equation}
n (s) \sim s^{-2}
\end{equation}
was proposed, where $n (s)$ refers to the number of sales of size $s$. This power-law is in agreement with our microscopic model, and our macroscopic model for $\alpha =1$ and $\beta = 1$. As mentioned in the work of Nagel {\it et al} \cite{nagel00}, an establishment is a single physical location at which business is conducted. A firm or company may consist of one establishment or more. Hence, our models should be in better agreement with firm distribution than establishment distribution because we do not introduce a spatial structure. However, there is no empirical evidence of any difference between both distributions, suggesting that the size distribution can only be weakly dependent on space. The models of ref. \cite{nagel00} are richer and more ambitious than ours, because they eventually would like to construct a model where the acceptance of money, the emergence of competitive price and the emergence of market structure all arise from the system dynamics \cite{nagel00}. Having taken a simpler approach, we are able to find analytical solutions to our models. Another interesting feature of our microscopic model is that it generates a time scale separation between bankruptcy and customer choice, a property that was introduced by hand in \cite{nagel00}.  

As in our models most businesses are proposing a similar price, the number of customers in one business is also a measure of the income of that business. As such, we can consider the company income distributions to see if our models do agree with the dynamics of companies. Okuyama {\it et al} \cite{takayasu99} considered the income distribution of companies in Japan. They proposed a Zipf 
plot of the incomes, ranking companies according to their income, and obtained a power-law of exponent of -1 for the income ranking plot, which is equivalent to a power-law of exponent -2 for the income distribution. This result for the aggregate data of different sectors companies is also obtained for most sectors taken separately. 

Aoyama {\it et al} \cite{takayasu00} investigated the debt of bankrupt companies in Japan, from 1997 to the end of March 2000. Their data are not conclusive but they suggest that a power-law of exponent -2 for the size distribution of debts is reasonable.

The results of ref. \cite{nagel00,takayasu99,takayasu00} suggest there is a universality in the size distribution of selling entities, with a power-law distribution of exponent close to -2 in the limit of large entities. This is in agreement with our models where such a power-law is found when companies are competing in a price driven market. However, there has been work on the size distribution of companies in different countries \cite{takayasu98,ramsden00}, that concluded that these distributions are not universal. Most of these distributions can be well approximated by a power-law of exponent close to -2, in Spain, Norway or Germany for instance. However, careful analysis suggests clear differences from one country to the other. Some countries do display a particularly different distribution, such as South Africa. It is difficult to clearly identify the origin of these distinctive features, but they suggest that our microscopic model is only able to generate the universal feature of the problem, which is exactly what we were aiming for. If one wants to generate a whole set of different distributions, it seems almost certain that a richer model is needed. We address this problem in a later work \cite{nato2001}.

A particular point of interest of our models is that they isolate $\beta = 1$ as a special value. As we mentioned earlier, the effective bankruptcy rate $a_{eff}$ scales like $a_{eff} \sim s^{1-\beta}$. So, our models predict that the bankruptcy rate is independent on the business or company size. This has indeed been observed in \cite{amaral97}. 

We should mention that in this study, we have primarily investigated the mean-field model with respect to its relation to the microscopic model. This particular choice leads us to assume that $a (s)$ is a power-law. However, taking $a (s) = (\ln s)^{\beta} / s$, we obtain that $n_s$ is a lognormal distribution for $\beta = 1$ \cite{nato2001}. This result is encouraging as most studies conclude that the size distribution of companies is a lognormal distribution for typical company sizes \cite{amaral97}, and a power-law for larger sizes\cite{nagel00,takayasu00}. Hence, the microscopic model we introduce offers a potential mechanism to generate a size distribution with a power-law in agreement with the results of Ref. \cite{nagel00}, but the mean-field model has a wider range of validity, depending on the particular functional form of $a (s)$. This functional form could even be country dependent, as variations from one country to the other have been empirically identified \cite{takayasu98,ramsden00}.    
 
\section{Conclusions}
\label{sec:conclusions}

To summarise, we have presented a microscopic model and an associated mean-field model to investigate the size distribution of selling organisations, that we choose to call businesses for simplicity. The microscopic model is based on customers trying to find the business that best matches their price expectation, while businesses have to comply with the market price to avoid bankruptcy. We showed numerically and analytically that such a dynamic allows only businesses selling at the market price to grow, cheaper businesses suffering financial slump while more expensive businesses cannot attract customers. The size distribution $n_s$ of the businesses, the size of a business being defined to be its number $s$ of customers, is a power-law $n_s \sim s^{-\tau}$ with an exponent $\tau$ close to 2. In the mean-field version of our model, the need to comply with the market price is replaced by a probability $a (s)$ that a business of size $s$ goes bankrupt. Taking $a (s) = \alpha s^{-\beta} n_1$, the asymptotic behaviour of $n_s$ is determined for every values of $\beta$. We showed that the microscopic model corresponds to $\beta$ equal to 1 and $\alpha$ close to but bigger than 1, going towards 1 as the number of customers goes to infinity. For $\beta = 1$, the exponent of the size distribution is equal to $\tau = 1 + \alpha$. The moments of the size distribution are determined, as well as the connectivity of the network of customers and the market price variations. The results for the microscopic model are in agreement with numerous empirical data, and we cited recent results about the size of companies annual sales in the U.S. \cite{nagel00} and about the size of debts of bankrupt companies in Japan \cite{takayasu00}. We also stressed that we have focussed our attention on an analytical form for $a (s)$ inspired by the microscopic model, while other forms for $a (s)$ can lead to a lognormal distribution for the size of the companies, an empirical result put forward by several authors \cite{amaral97}.


\begin{thebibliography}{99}
\bibitem{marks_and_spencer}
E. Robinson, Financial Times of the 16th of July 1999.

\bibitem{smith_corona}
J. Leffall, Financial Times of the 24th of May 2000.

\bibitem{boo_com}
C. Daniel, Financial Times of the 20th of May 2000.

\bibitem{dhulst00}
R. D'Hulst and G. J. Rodgers, Physica A {\bf 280}, 554 (2000) (adap-org/9912003).

\bibitem{dhulst-apfa2}
R. D'Hulst and G. J. Rodgers, proceedings of APFA 2, Li\`ege, July 2000, to be published in Eur. Phys. J. B.

\bibitem{pareto}
V. Pareto, {\it Cours d'\'economie politique} (Lausanne, 1897).

\bibitem{gibrat}
R. Gibrat, {\it Les In\'egalit\'es Politiques} (Sirey, Paris, 1931).

\bibitem{zipf}
G. K. Zipf, {\it Human behavior and the principle of least effort} (Addison-Wesley, London, 1944).

\bibitem{champerone}
D. G. Champernowne, The Economic Journal {\bf 63}, 318 (1953).

\bibitem{simon}
H. A. Simon, Biometrika {\bf 42}, 425 (1955).

\bibitem{simon2}
H. A. Simon and C. P. Bonini, The American Economic Review {\bf 48}, 607 (1958).

\bibitem{mandelbrot-levy}
B. Mandelbrot, Journal of Business {\bf 36}, 392-417 (1963).

\bibitem{steindl}
J. Steindl, {\it Random processes and the growth of firms} (Griffin, London, 1965), footnote of p. 18.

\bibitem{stanley-model}
S. V. Buldyrev, L. A. N. Amaral, S. V. Havlin, H. Leschhron, P. Maass, M. A. Salinger, H. E. Stanley and M. H. R. Stanley, J. Phys. I France {\bf 7}, 635 (1997) (cond-mat/9702085).

\bibitem{sid2}
S. Ipolatov, P. L. Krapivsky and S. Redner, Eur. Phys. J. B {\bf 2}, 267 (1998) (cond-mat/9708018).

\bibitem{takayasu98}
H. Takayasu and K. Okuyama, Fractals {\bf 6}, 67 (1998).

\bibitem{amaral97}
L. A. N. Amaral, S. V. Buldyrev, S. Havlin, H. Leschhorn, P. Maass, M. A. Salinger, H. E. Stanley and M. H. R. Stanley, J. Phys. I France {\bf 7}, 621 (1997) (cond-mat/9702082).

\bibitem{nato2001}
R. D'Hulst and G. J. Rodgers, proceedings of {\it Application of Physics in Economic Modelling}, Prague, February 2001, to be published in Physica A.

\bibitem{nagel00}
K. Nagel, M. Shubik, M. Paczuski and P. Bak, Physica A {\bf 287}, 546 (2000) (nlin/0005018).

\bibitem{sid_and_paul}
P. L. Krapivsky, F. Leyvraz and S. Redner, Phys. Rev. Lett. {\bf 85}, 4629 (2000) (cond-mat/0005139).

\bibitem{takayasu99}
K. Okuyama, M. Takayasu and H. Takayasu, Physica A {\bf 269}, 125 (1999).

\bibitem{takayasu00}
H. Aoyama, W. Souma, Y. Nagahara, M. P. Okazaki, H. Takayasu and M. Takayasu, Fractals {\bf 8}, 293 (2000) (cond-mat/0006038).

\bibitem{ramsden00}
J. J. Ramsden and Gy. Kiss-Haypal, Physica A {\bf 277}, 220 (2000).

\end{thebibliography}
\end{document}